\documentclass[ 10pt ]{article}

\usepackage[numbers,sort&compress]{natbib}
\usepackage[hyphens]{url}  
\usepackage{graphicx}
\usepackage{subcaption}
\usepackage{hyperref} 

\title{Detecting Anomalous Topology, Routing Policies, and Congested Interconnections at Internet Scale}
\author{Matt Mathis\thanks{Self supported independent researcher at Measurement Lab}}

\begin{document}
\maketitle

\begin{abstract}
Separating mid-path Internet performance from edge effects remains a fundamental challenge in network measurement. This paper presents a methodology for detecting anomalous topology, routing policies, and congested interconnections using controlled A/B comparisons derived from Measurement Lab (M-Lab) data. The approach leverages M-Lab's uniform server selection policy: by comparing performance distributions from clients in the same access ISP to different nearby M-Lab servers, natural experiments are created that isolate mid-path effects while controlling for client-side variation, access network bottlenecks, and diurnal variation in test volume. This analysis is implemented in BigQuery using sparse multidimensional histograms enabling efficient computation of Kolmogorov-Smirnov distance and ratios of geometric mean throughput across many millions of measurements in a single pass. Differences in throughput suggest mid-path bandwidth bottlenecks or traffic management; excess differences in minimum RTT suggest suboptimal routing.   These signals of interconnection problems are extracted from the noise deliberately suppressed by other measurement approaches.   Public dashboards provide ongoing visibility into all M-Lab metropolitan regions with sufficient servers, with drill-down capability to individual ISP--server plots.

\end{abstract}
\section{Introduction
}

This paper introduces a technique for detecting otherwise hidden mid-path problems, using data from Measurement Lab (M-Lab). The approach leverages M-Lab's uniform server selection policy to create natural A/B comparisons of performance measurements from users in the same access ISP to different nearby M-Lab servers. When all mid-paths perform similarly, one can be confident the measurements reflect genuine user experience as determined by the access ISP. When performance differs between servers, this indicates problems in the middle of the network such as insufficient capacity, congested interconnections, suboptimal routing, or missing physical connectivity.  The signals that other measurement approaches treat as noise become the primary signal.

The methodology relies on two key constraints to ensure valid comparisons. First, clients are grouped by their access ISP, as identified by their origin ASN (Autonomous System Number). This grouping isolates the performance characteristics of each access ISP's network and service tiers. Second, M-Lab's Locate Service uniformly distributes tests across geographically nearby M-Lab servers, explicitly disregarding network conditions, delays, or historical performance. This uniform test distribution, independent of network conditions, creates statistically equivalent measurement samples for each access ISP-server pair. Matching performance across different servers indicates clean mid-paths, while discrepancies signal potential anomalies.

This analysis is applied to each access ISP within a metropolitan region (an M-Lab metro), enabling evaluation of numerous paths (M × N, where M is the number of M-Lab servers and N is the number of ISPs with sufficient data).  This technique can be applied to multiple metrics collected by Linux \texttt{tcp\_info}. Large differences in throughput suggest mid-path bandwidth bottlenecks, while large differences in minimum Round-Trip Time (minRTT) suggest suboptimal topology, routing, or peering.

For detailed investigation, one examines fine-grained histograms and cumulative distribution function (CDF) plots of performance between individual access ISPs and multiple M-Lab servers. For global or regional overviews, these distributions are condensed into two difference statistics: the Kolmogorov-Smirnov (KS) distance and the ratio of geometric means.  These aggregated metrics are presented in global bar charts that allow rapid assessment of interconnection health across all monitored regions.

This technique is broadly applicable to M-Lab NDT7 data and potentially adaptable to other network measurement platforms, provided that server selection is not designed to mask interconnections nor influenced by observed performance nor other network properties, such as the RTT to the server.  
\section{Background
}

\subsection{Internet Interconnection Structure
}

The Internet is composed of thousands of independently operated networks, each identified by a unique AS Number (ASN). Access ISPs connect end users to the Internet; transit providers carry traffic between other networks; and cloud or content providers host compute and web services; although in practice most ISPs perform multiple roles simultaneously.  Traffic between an end user and a remote service typically crosses one or more \emph{interconnections}, points where independently operated networks exchange traffic.

The performance of these interconnections depends on their capacity, the routing policies of the networks involved, and the business relationships governing traffic exchange. When interconnection capacity is insufficient or routing is suboptimal, users experience degraded performance that cannot be attributed to their access ISP. Detecting and localizing these mid-path problems is the focus of this paper.
\subsection{Measurement Lab
}

Measurement Lab (M-Lab) is a consortium of research, industry, and public-interest partners that operates an open, verifiable measurement platform for global network performance~\cite{Dovrolis2010mlab,Gill2019mlab,Ohlsen2021advocacy,MLab2024about}.  By making data and methodology publicly available, M-Lab enables researchers, policymakers, ISPs, and consumers to better understand Internet infrastructure.
 \autoref{appendix} presents an extended history of the project and explains the design decisions, rationale and their impact on measurement.

M-Lab's primary measurement tool is the Network Diagnostic Tool (NDT),~\cite{Carlson2003NDT,carlson2003NDTcode}, which performs  single-stream TCP upload and download transfers between the user's device and an M-Lab server, recording throughput, minimum round-trip time (minRTT), and other TCP diagnostics via Linux's \texttt{tcp\_info} interface. All measurement data is archived and publicly available via Google BigQuery.

 By deploying servers in transit and content networks rather than within access networks, M-Lab explicitly measures the quality of interconnections as part of users' end-to-end experience.

Two architectural choices make M-Lab's data particularly suited to interconnection analysis:

\begin{itemize}
\item \textbf{Server placement in transit networks.} All early M-Lab servers were hosted in transit providers rather than within access ISPs.  As the Internet serving is evolving~\cite{Labovitz2010interdomain,Gigis2021hypergiants} M-Lab is now placing servers in other types of providers that are co-located with content, for example virtual servers in clouds, and selected access ISPs.    Servers were placed to maximize the number of  measurements that traverse one or two interconnections between the user's access network and the server's network~\cite{Gill2019mlab,Ohlsen2021advocacy}.  End-to-end measurement and the server placement strategy reflect architectural principles that have been with the project from the very beginning.  See \autoref{appendix} for more information about its history and rationale.

Measurement platforms that place servers on-net (within access networks) or select the topologically ``nearest'' server by network proximity or RTT tend to measure only the access edge, masking interconnection effects.

\item \textbf{Uniform server selection.} M-Lab's Locate Service assigns user’s tests to servers based solely on geographic proximity, without considering network conditions, latency, or historical performance. Within a metropolitan region, each nearby M-Lab server receives a statistically equivalent sample of user tests from each access ISP. This creates natural A/B comparisons: any systematic performance difference between servers serving the same user population must originate in the differing mid-paths, not in the access network or user equipment.
\end{itemize}

\subsection{Key Metrics
}

This analysis relies on two per-measurement metrics recorded by NDT:

\begin{itemize}
\item \textbf{Throughput} (Mean Download Throughput) reflects the end-to-end capacity available to a single TCP flow from the server to the client. Differences in throughput from different servers to the same clients suggest mid-path bandwidth constraints such as congested interconnections or traffic management (e.g. shaping).

\item \textbf{Minimum RTT} (minRTT) approximates the propagation delay of the end-to-end path. Excessive differences in minRTT between different servers as seen from clients on a single access ISP suggest suboptimal routing.  e.g., traffic ``hairpinning'' through a distant interconnection rather than peering locally.
\end{itemize}
There are several additional metrics that could be studied in the future, including loss rate and various measures of queueing or jitter.

\subsection{Difference Statistics
}

To summarize performance differences between paths, the following two complementary statistics are computed over pairs of measurement distributions:

\begin{itemize}
\item The \textbf{Kolmogorov-Smirnov (KS) distance}~\cite{wikiKSdistance} measures the maximum vertical separation between two cumulative distribution functions. It is sensitive to differences anywhere in the distributions, without any assumptions about the shapes of the distributions.  Although the KSdistance is robust, it lacks an intuitive interpretation for many readers.

\item The \textbf{ratio of geometric means}\footnote{The analyses in this study generally use geometric means and logarithmic X axis. Arithmetic means and linear graph axes don't make much sense because the data spans multiple orders of magnitude.},
termed ``spread’’ for brevity, provides an intuitive summary: one path is X\% faster (or slower) than another. However, the spread can obscure problems that affect only part of the distribution and may produce misleading results when distributions differ in offsetting ways across different ranges.
\end{itemize}

Small distance statistics confirm that mid-paths are not differentially affecting performance. Large distance statistics signal anomalies warranting further investigation.
\section{Examples
}
\begin{figure}
\centering
\begin{subfigure}{\columnwidth}
  \includegraphics[width=0.8\columnwidth]{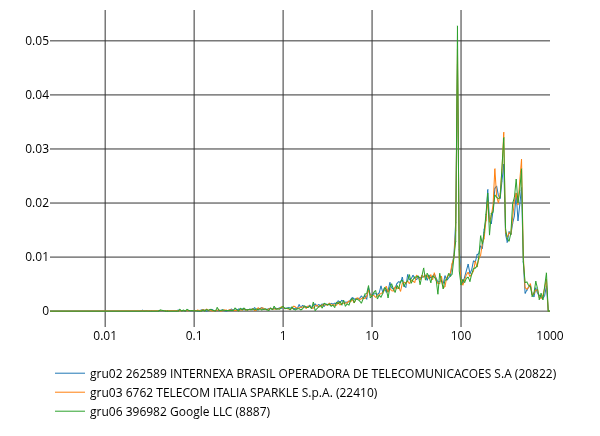}
  \caption{Download Throughput (Mbit/s) for Telefônica (AS 18881).}
  \label{fig:telefonica}
\end{subfigure}

\begin{subfigure}{\columnwidth}
  \includegraphics[width=0.8\columnwidth]{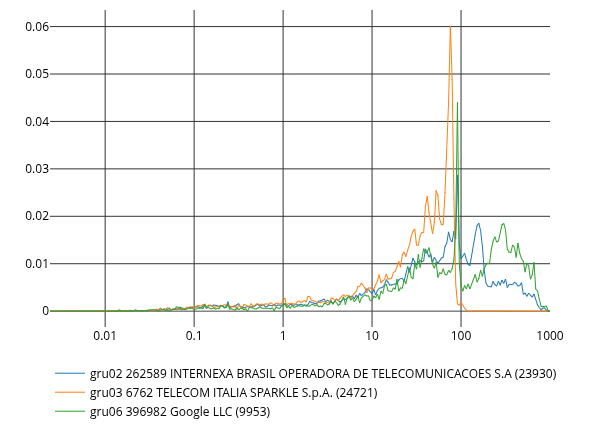}
  \caption{Download throughput for Claro (AS 28573).}
  \label{fig:claro}
\end{subfigure}

\caption{Two different access ISPs in São Paulo Brazil from the last week of November 2025.   There is no evidence of any mid-path bottlenecks circa Telefônica, and clear evidence of at least 2 mid-path bottlenecks circa Claro. 
The M-Lab servers are labeled by their ISP’s AS Number, name and the sample size.}

\label{fig:comparison}
\end{figure}

Figure~\ref{fig:comparison} illustrates the methodology using data from São Paulo, Brazil, where M-Lab operates three servers.   Figure~\ref{fig:telefonica} shows the probability density function (PDF) of download throughput for Telefônica, one of Brazil's largest ISPs and the legacy of the Brazilian PTT, as measured by tests to each of the M-Lab servers in São Paulo.   The close match between measurements from different servers confirms that neither the servers themselves nor the mid-paths significantly affect the results.  The tight agreement across measurement vantage points also suggests that the modes visible in Figure~\ref{fig:telefonica} accurately reflect Telefônica's service tiers or their users' WiFi or LAN bottlenecks.

Figure~\ref{fig:claro} shows a markedly different pattern for Claro S.A., another large Brazilian ISP. Performance to the three M-Lab servers varies substantially: gru03 (orange) appears to have a 100 Mb/s cap, while gru02 (blue) shows lower test populations than gru06 above 50 Mb/s.  These results indicate that mid-paths between Claro and at least two of the M-Lab servers substantially affect measured performance.  Although one might suspect that gru06 might reflect Claro’s service tiers, there is no corroborating evidence, and it is possible that all three M-Lab servers have different mid-path bottlenecks.  Figure~\ref{fig:telefonica} also demonstrates that the servers themselves are well calibrated, yielding identical results across their full throughput range; confirming that the results in Figure~\ref{fig:claro} were not caused by any differences in the servers.

\begin{figure}
  \centering
  \includegraphics[width=0.8\columnwidth]{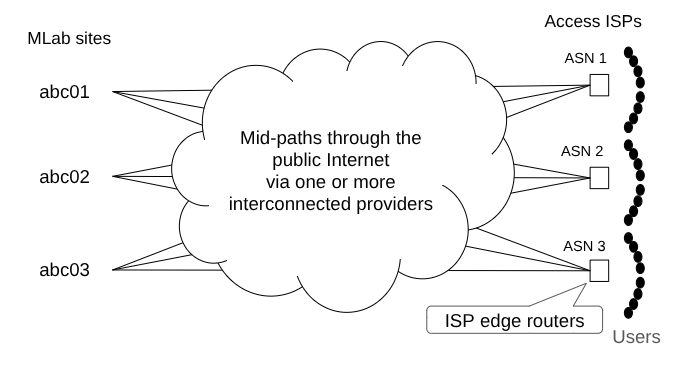}
  \caption{Mid-paths under test}
  \label{fig:model}
\end{figure}

Figure~\ref{fig:model} illustrates a conceptual model for attributing differences in Figure~\ref{fig:claro} to mid-path characteristics.  It shows three generic M-Lab servers on the left and three generic access ISPs on the right (such as Telefônica and Claro). In this example the servers and access ISPs are connected by nine different paths through the Internet.

The conceptual model uses a ``single attachment’’ assumption: each access ISP has one primary attachment to the global Internet through which all traffic passes.   This assumption is almost never strictly true in practice, however it is not required for the actual analysis.

Under the single attachment assumption, all end-to-end paths can be decomposed into two independent subpaths: a local path within the access ISP and its customers, and a wide-area path across the public Internet.  M-Lab's Locate Service distributes test traffic uniformly across all server-access ISP combinations within a region, ensuring that samples exercising each wide-area path have identical statistical properties,  determined by user’s test times and local bottlenecks. This design creates robust synthetic A/B comparisons between different paths across the public Internet..

When an access ISP has redundant connections, it potentially creates ambiguity in the distinction between mid-path and access bottlenecks.  This generally does not affect the overall analysis unless the access ISP has internal bottlenecks, in which case the ``mid-path’’ is effectively extended into the access ISP.
\section{Methodology
}
For each metric Multidimensional histograms are computed, with axes representing: M-Lab server, access ISP (by AS Number), and logarithmically-binned metric values (throughput, minimum RTT, loss rate, etc.). Each measurement is counted into exactly one bin per metric, requiring a single pass through M-Lab's data.   For efficiency the histograms are sparse:, bins with zero counts are not stored.  Figure~\ref{fig:comparison} shows PDFs  generated directly from these histograms.

For each metric and large access ISP, two difference statistics are computed for every pair of servers in the region: the ratio of their geometric means and their Kolmogorov-Smirnov (KS) distance. This typically includes the N top access ISPs and every pair of servers in every metropolitan region.  These difference statistics can be computed globally in a single query for all server pairs and relevant access ISPs.   By default the 5 largest access ISP in each metro are included, but that is entirely selectable.

Multiple dashboards summarize these statistics using different filters. For example, to verify server calibration, dashboards confirm that every server is part of at least one pair of servers with small difference statistics within its region.

\begin{figure}
  \centering
  \includegraphics[width=0.8\columnwidth]{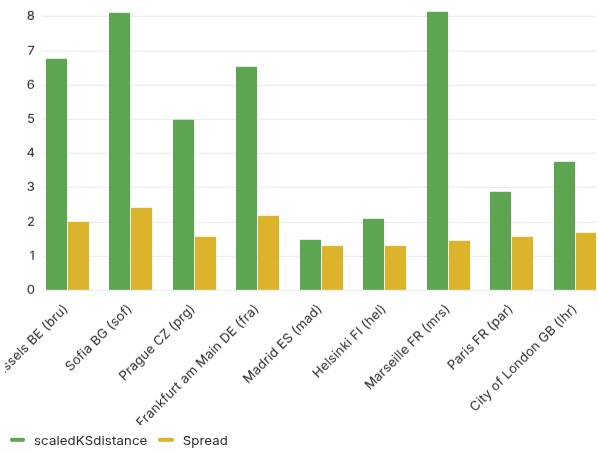}
  \caption{Worst case minRTT difference statistics in each of 10 European metros.  The vertical axis is a ``pain’’ metric, either 10 times the KS distance or the ratio of the minRTTs (``spread’’).  The full live bar chart~\cite{MLab2025barchart} shows 50 metros having at least 2 M-Lab servers.  You can click through any bar to access detailed performance plots for the metro.  The M-Lab fleet includes another 60 metros that will show difference statistics as we deploy additional servers.}
  \label{fig:bar}
\end{figure}

To assess interconnection health, the server pairs with the largest difference statistics are identified within each metro region.  Figure~\ref{fig:bar} shows a representative fragment of a bar chart displaying the maximum difference statistics in each metro where M-Lab operates multiple servers. Large differences indicate that users in different access ISPs experience varying performance to the content depending on which transit providers carry the traffic. Sufficiently large differences may indicate that some users cannot effectively access some nearby content, suggesting an unhealthy Internet.

The complete current bar chart is available as a public interactive dashboard at~\cite{MLab2025barchart}. This dashboard has separate global bar charts for showing the worst case differences in minRTT and MeanThroughput.   The public dashboards are updated weekly.

The code has seriously outgrown the development tools and plot scaffolding (BigQuery Studio and grafana).    As a consequence this paper only covers two of the most important dashboards.   There are several unfinished running prototypes of other types of analysis, including differential time series, other ways of comparing mid-paths and investigation of additional TCP parameters.  Every new dashboard raises more questions than it answers and leads to additional prototypes.  The queries will be published on GitHub as they are moved to new tools chains and better scaffolding.
\subsection{Interpretation  
}

MinRTT typically reflects the path length. Excess differences in minimum RTT suggest suboptimal routing, for instance, paths between ISPs that lack direct local interconnection and instead route traffic through remote interconnections.   Even when the long routes do not reduce throughput, they degrade application responsiveness and increase page load times~\cite{Cheshire1996latency,Gettys2012bufferbloat,Briscoe2016reducing,BITAG2022latency}.  In heavily congested networks, minimum RTT may also include persistent queuing delays, indicating more severe problems than just routing inefficiency alone.

The second bar chart on the live dashboard shows maximum throughput differences in each metro. This identifies situations where connectivity exists but mid-path capacity or traffic management limits end-to-end throughput. The most common case is insufficient network capacity, which typically affects fast flows more than slow flows.   The differences between gru02 and gru06 in Figure~\ref{fig:claro} suggest a congested interconnection.

In comparison, the sharp upper bound and narrow 100 Mb/s mode for gru03 strongly suggest a stateful per-flow policer or rate limiter somewhere in the end-to-end path.  If the bottleneck was applied to aggregate traffic, the cross traffic from multiple streams would smear the performance mode of each test.  The sharpness of the 100 Mb/s mode implies that each individual test flow encounters its own bottleneck, rather than sharing aggregate capacity with other flows.

The live bar charts~\cite{MLab2025barchart} provide drill-down capability to metro-specific dashboards containing full suites of detailed performance plots such as Figure~\ref{fig:comparison}.   To access them click on the bar of interest to make the tool tip ``sticky’’ and then click on the link at the bottom of the tooltip.   There are thousands of detailed plots just a few clicks away from the bar chart.

For every data point in the plots M-Lab has full \texttt{tcp\_info} traces, a majority have concurrent scamper traceroutes~\cite{Luckie2010scamper,CAIDA2025scamper} and about 10\% have full pcap packet captures.   The data could be used to augment other methods of investigating mid-path congestion such as TSLP\cite{Sundaresan2017tcp,Dhamdhere2018inferring} and MANIC\cite{CAIDA2020manic}.
The differential analysis might also be aligned with BGP RouteView to infer opportunities for better peering.  It could also be used to search for interesting parts of the Internet to probe with other tools and techniques.
\section{Conclusion
}
This paper presents a methodology for detecting mid-path network anomalies at Internet scale. Key contributions and findings include:
\begin{itemize}

\item Controlled A/B comparisons enabled by M-Lab's uniform server selection policy isolate mid-path effects while controlling for access network bottlenecks, client-side variation, and diurnal variation in test volume.

\item Variation between mid-paths are the primary signal for detecting interconnection problems.

\item These are the same signals that other techniques strive to exclude as measurement noise.

\item Throughput differences between servers identify congested interconnections and mid-path traffic management. MinRTT differences reveal suboptimal routing and hairpinning, degrading application responsiveness even if throughput is unaffected.

\item Sparse multidimensional histograms enable efficient global-scale analysis. A single query computes distance metrics (KS distance and ratio of geometric means) for all combinations of server pairs and access ISPs with sufficient data.

\item Public dashboards provide ongoing monitoring of interconnection health across all M-Lab metropolitan regions, with drill-down capability to individual ISP--server plots for operators, researchers, and policymakers.

\item The code used in this study will be ported to a smarter tool chain and scaffolding to improve the scaling of the software development itself.   Multiple existing prototype dashboards will be finished and published.

\item As the underlying SQL queries are moved to GitHub they will be published to facilitate replication and collaboration.

\item Research opportunities include using M-Labs data to augment other interconnection and topology research such as integrating scamper/traceroute data, alignment with RouteView BGP data, and identifying paths exhibiting anomalous behavior for deeper investigation.  M-Lab probably would not lead these efforts, but would collaborate with other teams.

\item Insights from these dashboards will help ISPs incrementally anneal the global Internet, one peering relationship at a time.

\end{itemize}

\bibliographystyle{unsrtnat}
\bibliography{refs}

\appendix
\section{The Back Story
}
\label{appendix}

Some of the material in this section reflects the author's direct involvement, firsthand conversations and interpretations, many of which are not supported by citable documents.

Historically M-Lab deployed servers in transit networks, such that M-Lab explicitly measured network interconnections between transit and access ISPs as part of users' end-to-end experience.~\cite{Dovrolis2010mlab,Gill2019mlab,Ohlsen2021advocacy,MLab2024about}   Due to changes in the Internet itself and the shift to hypergiants~\cite{Labovitz2010interdomain,Gigis2021hypergiants}, M-Lab has been diversifying its fleet to include cloud providers, selected access ISPs and other serving platforms.   The vast majority of the tests still traverse a few interconnections between the server's ISP and the user’s access ISP.

\subsection{NDT History
}

M-Lab and its most important tool, NDT, were both inspired by earlier work in the IETF IPPM working group~\cite{IPPM} and specifically RFC~2330~\cite{RFC2330}, the working group’s foundational document.  This work explicitly considers Internet paths to be the concatenation of subpaths and imagines a future Analytical Framework that might represent end-to-end path properties as the spatial concatenation of the properties of its subpaths.

Early in his career the author had multiple painful experiences debugging network applications which worked well across university campus LANs, but could not run at acceptable performance when the path was extended with an underloaded lossless, effectively perfect, US scale backbone  network.  Short RTT flows across a campus could compensate for many network problems that were complete blockers over the longer path.  This phenomenon was called ``symptom scaling’’, because the symptoms of many network problems scale with the RTT.  Network flaws that cause only minor performance reduction over a campus LAN would render an application unusable when extended cross country.  A corollary is that testing a short subpath is not a good predictor of the performance as part of a longer end-to-end path, even if the rest of the path is perfect.   The author spent a lot of his early career developing tools to diagnose long paths~\cite{Mathis1994mping,Mathis1996TReno,Mathis2008NPAD}.

\subsection{The Interconnection Report
}

In 2014, the M-Lab team published ``ISP Interconnection and its Impact on Consumer Internet Performance’’\cite{MLab2014interconnection}, a landmark study utilizing M-Lab network measurement data across the United States.  The research focused on understanding how the interconnections between major access ISPs (such as AT\&T, Comcast, and Verizon) and transit ISPs (such as Cogent and Level 3) affected the  performance experienced by end-users. The report revealed a clear correlation between these interconnections and degraded consumer experience, which was most pronounced during peak usage hours, suggesting insufficient capacity and congestion as key contributing factors. Notably, the study suggested that these performance issues were tied to the business relationships between the ISPs rather than any  fundamental technical problem, as similar patterns of degradation were observed for specific pairs of ISPs across widely separated geographic locations sharing absolutely no common infrastructure.  The full story was quite complicated and included some surprises\cite{Rayburn2014cogent}.


This research provided the first large-scale, data-driven evidence of interconnection bottlenecks and the harm to the consumer Internet experience. The study informed public discourse and policy debates surrounding net neutrality, ISP responsibilities and FCC regulatory scope.  The subsequent FCC Open Internet Order~\cite{FCC2015open} reclassified broadband Internet access as a telecommunications service under Title II of the Communications Act and included  language that asserted the FCC's authority to hear complaints about interconnection arrangements on a case-by-case if they were deemed not ``just and reasonable.’’

In 2014 M-Lab was able to produce the interconnection report because some of the interconnections were so badly congested that the problems could be detected using relatively unsophisticated statistical techniques.   The huge diurnal variations in Internet performance were easily observed without compensating for measurement biases caused by M-Lab's self selected data samples and the diurnal variation in test volume.   M-Lab's detractors (primarily the ISPs themselves) correctly pointed out that ``people testing in anger’’ had the potential to magnify the observed performance variations, because they might test more when the network was performing poorly.

Following the 2015 FCC Open Internet Order, the large diurnal variations in Internet performance in North America diminished substantially~\cite{MLab2017monitoring,fcc2017mlabfiling}, to the point where diurnal variation in Internet performance were masked by the diurnal variation in M-Lab’s measurement volume. Multiple research efforts by the author and others between 2015 and 2024 attempted to address this ``signal searcher problem,’’ but no successful methods have been published to date.

The algorithm presented in this paper solves the signal searcher problem because the A/B comparisons are between two sub samples that have identical measurement statistics, including diurnal test volumes.
\subsection{ISP and FCC responses
}
The ISPs who were present at the creation of the IPPM working group were extremely unhappy with the developments there (firsthand conversations), and the idea that they might be evaluated on the performance of end-to-end paths, where they did not own and could not manage subpaths beyond their edges.   This position is completely understandable.  However, alternatives do not address symptom scaling, and the reality that measurements over short paths often fail to predict performance when extended with ideal networks to longer paths.

Although it was not recognized at the time, today it is understood that ISPs do have implicit control over the performance of their network peers, by the choices they make in selecting their peers.  In parts of the world where the aggregate internet capacity is large enough there will always be potential peers with enough headroom (available capacity) to carry consumer grade traffic without any degradation.   Today routing and interconnections in North America and Europe are reasonably healthy.  This is not true for most of the global south and pacific rim, where there is ample evidence of interconnection problems, even with M-Lab’s incomplete coverage.

When the ``Measuring Broadband America’’ (MBA) program was announced in 2010~\cite{FCC2010broadband,wiki2025broadband} it did not specify methodology nor make any mention of interconnections. It did focus on the most visible part of the Internet, the access edge, where customers (businesses and users) connect to ISPs.   During the Net Neutrality debates, there was a very contentious  conversation about the inclusion  of  interconnection measurements in the MBA program.  The FCC came to agree with the ISPs and the MBA program explicitly excluded interconnections.

This in turn, focused the research community on measuring the Internet access edge. Various commercial platforms (Ookla~\cite{Ookla}, SamKnows~\cite{SamKnows}), and academic studies adopted measurement methodologies that masked any performance variations that might have been caused by interconnections.   For example by preferring on-net servers; selecting servers based on network metrics such as RTT or topology; or aggregating concurrent tests to multiple servers, such that no individual server with poor performance would affect the aggregate performance.  While these approaches improved measurements of the access edge, by design they masked interconnection problems.  Comparative studies of measurement platforms~\cite{Goga2012speed,Bauer2016gigabit,Feamster2020challenges,MacMillan2023comparative,Lipphardt2025speed} all noted that M-Lab’s NDT generally showed lower performance and greater variance than other tools.  Many of the comparison studies attributed this to mid-path effects that other methodologies suppressed.

All of these comparison studies failed to note that M-Lab’s measurement technique explicitly covered  end-to-end measurement from off-net vantage points.

\subsection{Closing
}

One way to view this methodology is that the measurement traffic from each access ISP is used to probe the paths to all nearby M-Lab servers, without regard to the access ISP’s performance.  If the results are the same to all servers, then the access ISPs performance distribution can be readily observed from any of the nearby servers. If the performance differs then there must be bottlenecks in the mid-path, and one can not draw strong conclusions about the access ISP’s true performance distributions.

This methodology separates access edge performance from interconnection performance which is the very signal that other methods reject as noise.

\end{document}